\documentclass[12pt,a4paper]{article}
\usepackage[T2A]{fontenc}
\usepackage[utf8]{inputenc}
\usepackage[english]{babel}
\usepackage{amssymb}
\usepackage{amsmath}
\usepackage{graphicx}
\usepackage[small,sc,center]{titlesec}
\usepackage{indentfirst}
\usepackage{siunitx}
\usepackage[labelsep=period]{caption}
\usepackage{caption}

\newcommand{\nc}{\newcommand}
\nc{\etab}{\eta_\mathrm{B}}
\nc{\etar}{\eta_\mathrm{R}}
\nc{\lhe}{L_\mathrm{3\alpha}}
\nc{\mh}{M_\mathrm{H}}
\nc{\mzams}{M_\textrm{ZAMS}}
\nc{\pmin}{\Pi_{\min}}
\nc{\teff}{T_\textrm{eff}}
\nc{\tev}{t_\textrm{ev}}
\begin{document}

\begin{center}
	\textbf{The nature of anomalous period increase in the pulsating variable V725 Sgr}
	
	\vskip 3mm
	\textbf{Yu. A. Fadeyev\footnote{E--mail: fadeyev@inasan.ru}}
	
	\textit{Institute of Astronomy, Russian Academy of Sciences,
		Pyatnitskaya ul. 48, Moscow, 119017 Russia} \\
	
	Received October 19, 2022; accepted October 26, 2022
\end{center}

\textbf{Abstract} ---
Evolutionary tracks of stars with masses on the main sequence $0.84M_\odot\le\mzams\le 0.95M_\odot$
and initial metal abundances $Z=0.006$ and $Z=0.01$ were computed under various assumptions about
the mass loss rate at the red giant stage as well as at the AGB and the post--AGB stages.
Among 160 evolutionary sequences we selected nearly 30 sequences where the final thermal flash of the
helium shell source occurs in the early post--AGB stage when the ratio of the hydrogen envelope mass
to the stellar mass ranges from 0.01 to 0.08.
Selected evolutionary sequences were used for calculation of initial and inner boundary conditions
used in solution of the equations of radiation hydrodynamics and turbulent convection describing
evolution of stellar pulsations after the helium flash.
Among about three dozen hydrodynamic models we found the three ones demonstrating almost eightfold
increase of the pulsation period observed in V725 Sgr during the last century as well as the gradual
transformation of fairly regular pulsations with period $\Pi\approx 12$ day to semi--regular
non--linear oscillations with period $80~\textrm{day} \lesssim\Pi\lesssim 90~\textrm{day}$.
We conclude that the anomalous growth of the pulsation period in V725 Sgr is due to the final
thermal flash of the helium shell source that occured in the early post--AGB star with mass
$M\approx 0.53M_\odot$ and the mass of the hydrogen envelope ranging from $0.013M_\odot$ to
$0.019M_\odot$.


\newpage
\section*{introduction}

The variable star HV 7642 was discovered by Swope (1937) and at present is known as V725 Sgr
(Samus' et al. 2017).
Swope (1937) noted that the light curve of this variable resembles those of Cepheids but
at the same time it is remarkably different from them due to the unusually high rate of period
change.
In particular, during almost ten years from 1926 to 1936 the pulsation period of V725 Sgr
increased from 12 day to 21 day (Swope 1937).
The following photometric observations of this variable star were carried out in 1968 and
1969 by Demers (1973).
He showed that V725 Sgr belongs to population II stars but should be classified as
RV Tau or semi--regular type pulsating variable besause of the significantly increased
pulsation period ($45~\textrm{day}\lesssim\Pi\lesssim 50~\textrm{day}$).
The fact that V725 Sgr is the population II star was corroborated by Harris and Wallerstein (1984)
who investigated the kinematic properties of type II cepheids.
In 1973 the pulsation period of V725 Sgr was nearly 50 day (Demers and Madore 1974) and now is
as high as $\Pi\approx 90$ day whereas the light variations became less regular (Percy 2020).
Therefore, for the last century the pulsation period of V725 Sgr increased almost by a factor of eight
so that this variable gradually transformed from the population II cepheid into the long--period
semi--regular pulsating variable.

Percy et al. (2006) supposed that increase of the pulsation period observed in V725 Sgr is due to
the thermal flash in the helium shell source.
This assumption is based on comparison of the characteristic time of period change in V725 Sgr
with theoretically computed rates of radius changes in AGB stars undergoing helim flashes
(Vassiliadis and Wood 1993).
Population II cepheids are the low--mass post--AGB stars so that the loop of the evolutionary track
on the Hertzsprung--Russel (HR) diagram can cross the pulsation instability strip depending on
the hydrogen envelope mass during the final thermal flash  (Fadeyev 2020).
Pulsation period increase observed in V725 Sgr is a unique phenomenon for population II cepheids and
is of great interest to determine the fundamental parameters of the star by the methods of the
stellar evolution and nonlinear stellar pulsation theories.

The goal of the present study is that to verify the assumption proposed in Percy et al. (2006)
on the basis of consistent evolutionary and stellar pulsation computations and to reproduce the
period change observed in V725 Sgr.
To solve this problem we computed the hydrodynamic models using the time--dependent inner boundary
conditions describing evolutionary changes of the radius and the luminosity at the bottom of the
stellar envelope model.
In our preceding paper (Fadeyev 2022) we showed that in the framework of this approach the solution
of the equations of radiation hydrodynamics and turbulent convection is completely consistent with
results of stellar evolution calculations.
It should be noted, however, that the necessary condition for the existence of the looped evoltionary
track is that the thermal flash of the helium shell source should happen within the relatively short
time interval (a few tens of thousands of the years) when the mass of the hydrogen envelope decreases
from $\approx 8$ percent to $\approx 1$ percent of the stellar mass.
The thermal flashes are not strictly periodic and for the average interflash interval
$\langle\Delta t_\mathrm{tp}\rangle\sim 2\times 10^5$ yr the time interval between adjacent flashes
varies by nearly twenty percent.
Therefore, not all the loops of evolutionary tracks may lead to observed changes in the pulsation
period.
Due to this fact in the present work we considered almost 160 theoretically computed evolutionary
sequences of AGB and post--AGB stars.
However, among the variety of initial conditions we succeeded to obtain only a few hydrodynamic models
with period change which agrees with that observed in V725 Sgr.

\section*{evolutionary sequences of low--mass post--agb stars}

Initial conditions required to compute the non--linear stellar oscillations were determined from
evolutionary computations for stars with masses on the main sequence
$0.84M_\odot\le\mzams\le 0.95M_\odot$.
Metallicity of type II cepheids observed among field stars varies in a wide range and shows no strong
dependence on the galactocentric distance (Harris 1981).
Therefore, calculations of stellar evolution were carried out with two initial metal abundances:
$Z=0.006$ and $Z=0.01$ whereas the initial abundance of helim was assumed to be $Y=0.28$.

To calculate the evolutionary sequences we employed the program MESA version r15140 (Paxton et al. 2019).
Computational details of nucleosynthesis and convective mixing are discussed in our previous papers
(Fadeyev 2020, 2021).
Because the mass loss rates are very uncertain the evolutionary computations were carried out with
various assumptions about the mass loss rate $\dot M$ on the stage preceding AGB (Reimers 1975) as well as
on the AGB stage (Bl\"ocker 1995).
In particular, the Reimers mass loss rate formula was used with two values of the parameter
($\etar=0.3$ and $\etar=0.5$) whereas evolution on the AGB stage were computed with eight values of the
parameter $\etab$ ranged within $0.03\le\etab\le 0.1$ with step $\Delta\etab=0.01$.
In general we computed nearly 160 evolutionary sequences for the whole AGB and the early post--AGB
stages.

\begin{figure}
\centerline{\includegraphics{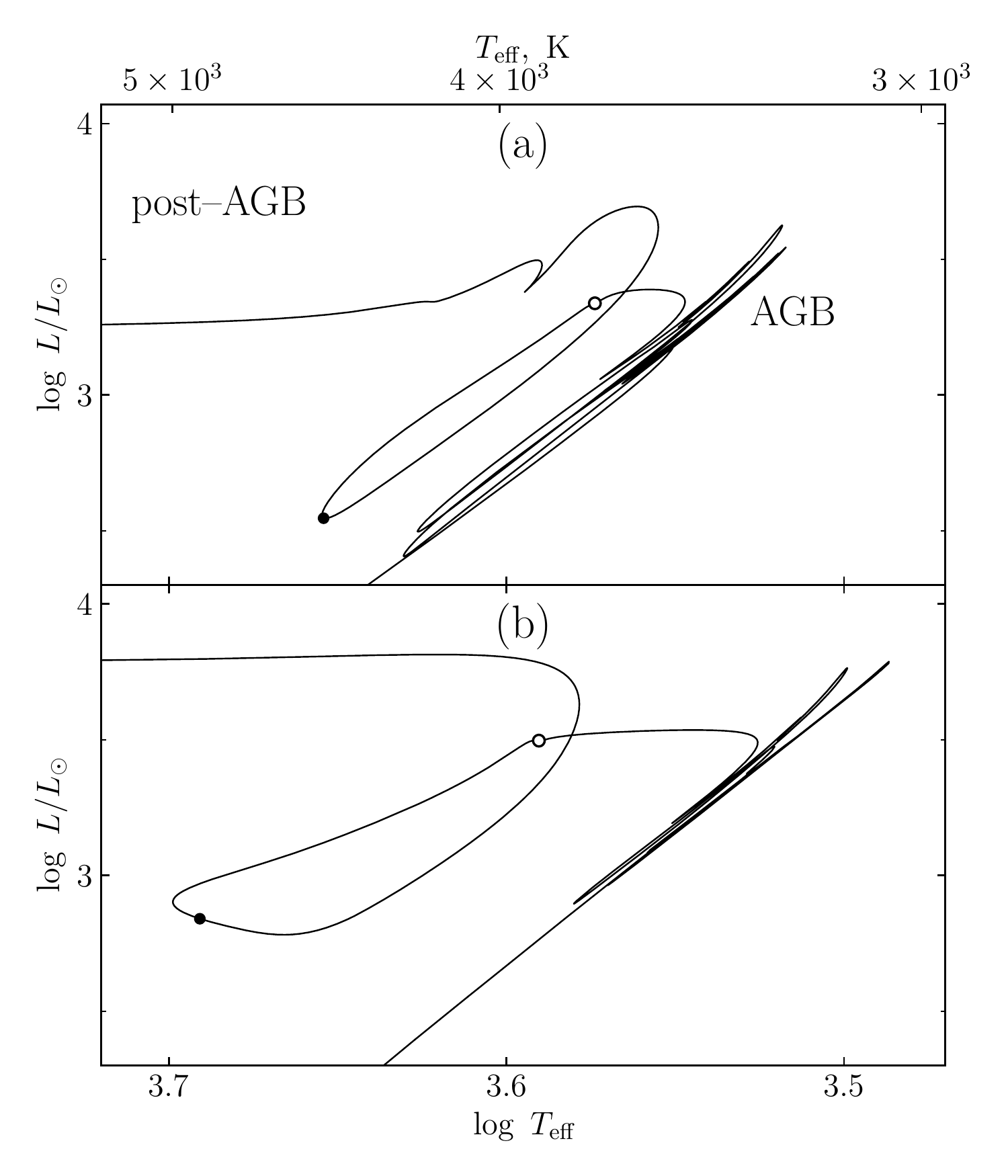}}
\caption{Tracks of evolutionary sequnces $\mzams=0.88M_\odot$, $\etar=0.5$, $\etab=0.08$, $\mh/M=0.025$ (a)
         and $\mzams=0.92M_\odot$, $\etar=0.3$, $\etab=0.07$, $\mh/M=0.013$ (b) for $Z=0.006$ during the AGB
         and the early post--AGB stages. Open circles correspond to the maxima of $\lhe$ and filled circles
         indicate the minimum radius of the star after the final helium flash.}
\label{fig1}
\end{figure}

Typical masses of stars at the end of the AGB stage are in the range $0.52M_\odot\le M\le 0.59M_\odot$
whereas the rapidly decreasing mass of the hydrogen envelope $\mh$ is nearly one percent of the
stellar mass.
If the final thermal flash occurs when the ratio of the hygrogen envelope mass to the stellar mas is
$\mh/M > 0.1$ the evolution of the post AGB star proceeds without loops at nearly constant luminosity
between the red giant and the high--temperature region of the HR diagram.
However, the post--AGB evolution is remarkably different if the final thermal flash occurs at smaller
ratios $\mh/M$.
Fig.~\ref{fig1} shows the tracks of two evolutionary sequences
$\mzams=0.88M_\odot$, $\etar=0.5$, $\etab=0.08$ and $\mzams=0.92M_\odot$, $\etar=0.3$, $\etab=0.07$
where the final thermal flash occurs for the ratios $\mh/M=0.025$ and $\mh/M=0.013$, respectively.
The points of the evolutionary tracks corresponding to the maximum luminosity of the helium shell source
$L_{3\alpha}$ are marked by the open circles.

As seen in Fig.~\ref{fig1}, decrease of the ratio $\mh/M$ at maximum  $L_{3\alpha}$ is
accompanied by displacement of the loop to higher effective temperatures.
Of main interest in computation of hydrodynamic models of V725 Sgr is the point on the loop
corresponding to the minimum stellar radius.
In Fig.~\ref{fig1} these points are marked by filled circles.
The period of radial pulsations is related to the stellar radius by $\Pi\propto R^{3/2}$
therefore the pulsation period reaches the minimum in the point of the minimal radius and then
gradually increases during the following evolution.
The main goal of the present study is to determine the theoretical dependence of the period of
radial oscillations as a function of the star age $\Pi(\tev)$ therefore the necessary condition
is that the star with minimal radius should reside within the instability strip.

\begin{figure}
\centerline{\includegraphics{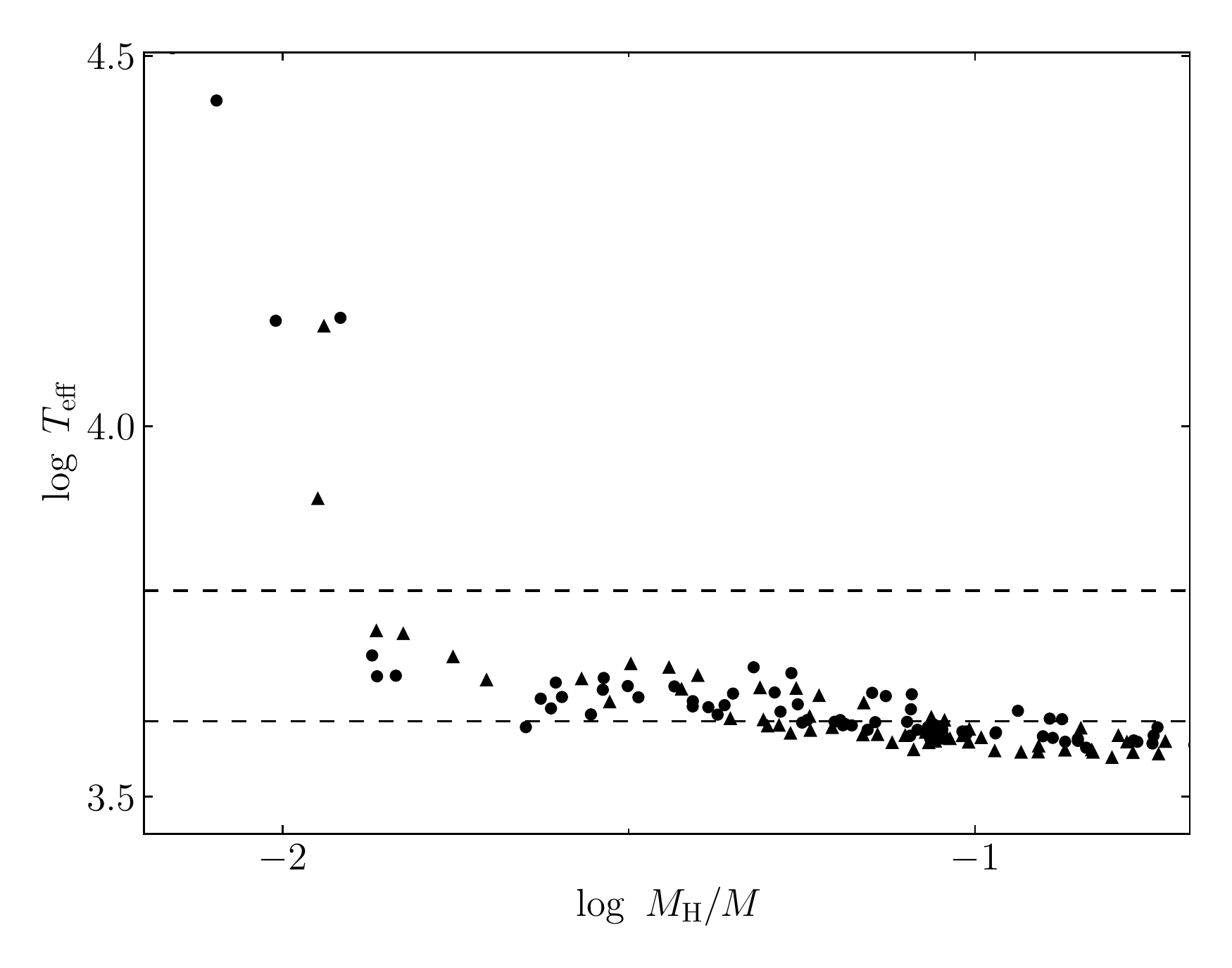}}
\caption{The effective temperature of the star with minimal radius versus the ratio of the hydrogen envelope to
         the stellar mass $\mh/M$. The circles and the triangles correspond to the evolutionary sequences
         computed with $Z=0.006$ and $Z=0.01$, respectively.
         Dashed lines indicate effective temperatures at the red ($\teff=4\times 10^3\:\mathrm{K}$) and
         at the blue ($\teff=6\times 10^3\:\mathrm{K}$) edges of the instability strip.}
\label{fig2}
\end{figure}

To determine the conditions necessary for location of the model with minimal radius after
the thermal flash within the instability strip we used the diagram shown in Fig.~\ref{fig2}, where
the effective temperature is plotted versus the ratio of the hydrogen envelope mass to the stellar
mass $\mh/M$.
For the sake of graphical clarity we present the plots for nearly 130 models with effective
temperatures $\teff < 4.5\times 10^4\: K$.
As seen in Fig.~\ref{fig2}, effective temperatures of models with minimal stellar radius do not show
dependence on metallicity $Z$ (the plots correspondong to evolutionary sequnces with $Z=0.006$ and
$Z=0.01$ are shown by circles and triangles, respectively) and depend only on $\mh/M$.

Calculations of non--linear pulsations of population II cepheids carried out by the author earlier
(Fadeyev 2020) as well as in the present study allow us to conclude that the edges of pulsation
instability nearly correspond to effective temperatures $\teff\approx 4\times 10^3\:K$ (the red edge)
and $\teff\approx 6\times 10^3\:K$ (the blue edge).
These estimates of $\teff$ agree with empirical results obtained by Demers and Harris (1974) and
are shown in Fig.~\ref{fig2} by dashed lines.

Therefore, to compute the hydrodynamic model we have to use the evolutionary sequnces where the
final thermal flash occurs for $0.02\lesssim \mh/M \lesssim 0.08$.
It should be noted that these values of thre ratio $\mh/M$ provide with one of necessary conditions
of applicability of the evolutionary sequence.
Verification of another condition assuming that the pulsation period of the star with minimal radius
is $\Pi\le 12$ day can be obtained only from trial calculations.
Moreover, evolutionary increase of the stellar radius after the radius minimum is accompanied by
decrease of the effective temperature (see Fig.~\ref{fig1}) so that many models locating near the lower
(red) edge in Fig.~\ref{fig2} were found to be inapplicable for modelling of V725 Sgr since they show
decaying oscillations after crossing of the red edge of the instability strip.

\section*{hydrodynamic models of v725 sgr}

Solution of the equations of radiation hydrodynamics and time--dependent convection describing
radial stellar oscillations (Fadeyev 2013) was obtained on the finite difference grid consisting
of 600 Lagrangian mass zones.
500 outer mass intervals increase inward from the upper boundary to the stellar center
geometrically whereas 100 inner intervals reduce with another value of the common ratio.
Such a distribution of Lagrangian mass zones allowed us to avoid the large approximation errors
in the inner layers of the pulsatring envelope where the gradients of pressure and temperature
sharply increase.
The inner boundary of hydrodynamic models is set in the layers with the gas temperature
$T\sim 5\times 10^6\:K$ and the radius $r_0\sim 10^{-2}R$, where $R$ is the radius of the upper
boundary of the evolutionary model.
The boundary separating the regions with different behaviour of mass interval locates in the layers
with temperature $T\sim 5\times 10^5\:K$ where the mass of outer layers is $\approx 2/3$ of the
stellar envelope mass.

During the helium flash the structure of the stellar envelope changes in the thermal time scale
so that the solution of the Cauchy problem for equations of hydrodynamics was obtained with
time--dependent inner boundary conditions explicitly describing temporal variations of the radius
and the luminosity: $r_0(\tev)$ and $L_0(\tev)$.
These dependences were determined from evolutionary computations for the fixed value of the
Lagrangian coordinate whereas the continuous functions $r_0(\tev)$ and $L_0(\tev)$ needed in
hydrodynamic computations were calculated using the cubic interpolating splines.
This method was earlier employed for explanations of abrupt decrease of pulsation amplitude in RU Cam
(Fadeyev 2021) and for modelling of the Mira--type pulsating variable T UMi undergoing the
thermal flash in the helium shell source (Fadeyev 2022).

\begin{figure}
\centerline{\includegraphics{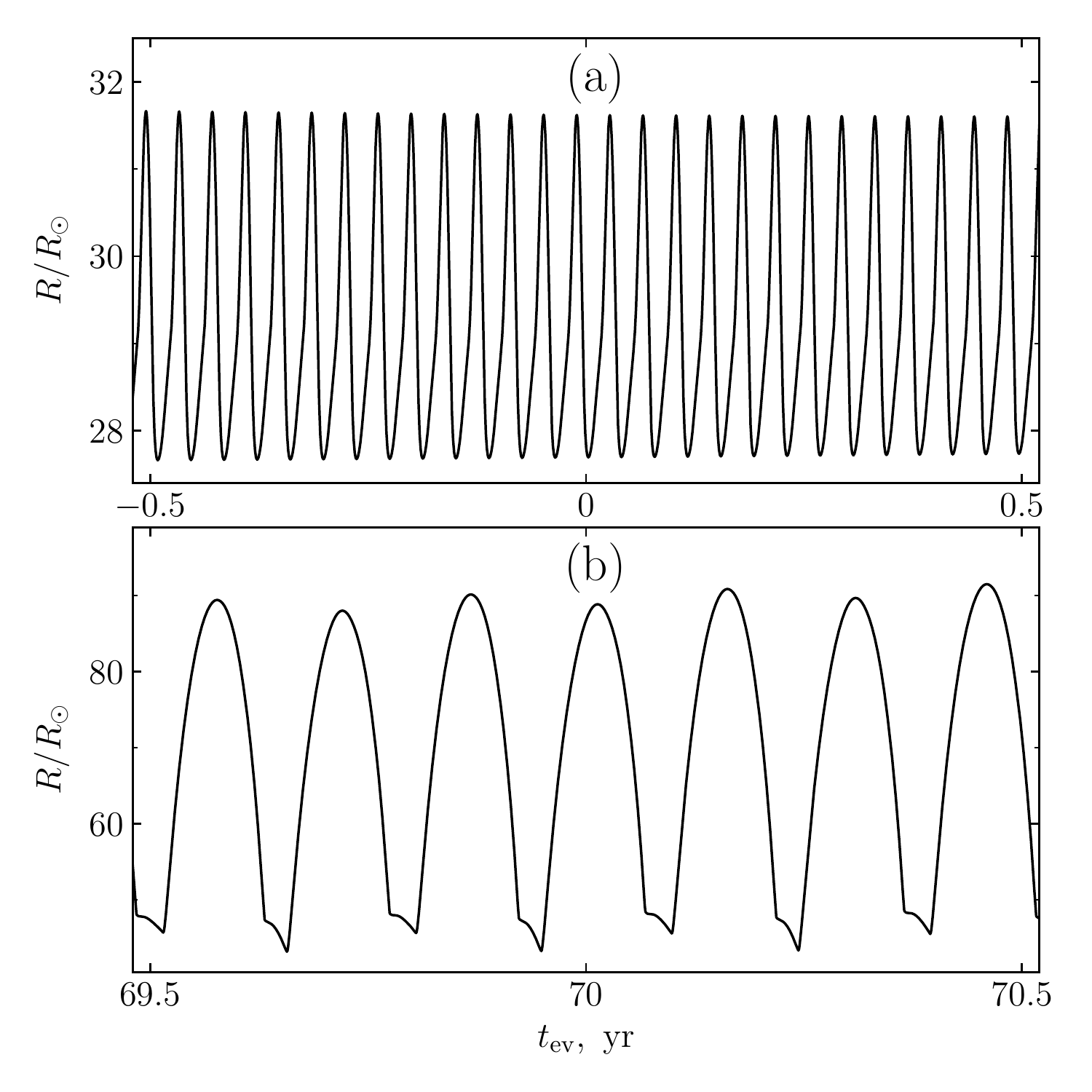}}
\caption{Variation with time of the upper boundary radius of the hydrodynamic model $Z=0.01$, $\mzams=0.9M_\odot$,
         $\etar=0.5$, $\etab=0.03$ in vicinity of the minimum radius of the evolutionary model $\tev=0$ (a) and
         after 70 yr (b).}
\label{fig3}
\end{figure}

Evolution of stellar pulsations after the helium flash is illustrated in Fig.~\ref{fig3} by the plots
of the upper boundary radius of the hydrodynamic model $Z=0.01$, $\mzams=0.9M_\odot$, $\etar=0.5$,
$\etab=0.03$.
For a better graphical representation the evolutionary time $\tev$ is set to zero at the minimum
radius of the evolutionary model when the pulsation period is $\Pi=13.9$ d.
The peak of the final helium flash corresponds to $\tev=-302$ yr.

As seen in Fig.~\ref{fig3}, in vicinity of the minimum radius the stellar pulsations are
characterized by sufficiently small amplitude of the radial displacement:
$\Delta R/\langle R\rangle\approx 0.12$, where $\langle R\rangle$ is the average radius of
the upper boundary of the hydrodynamic model.
After 70 yr the relative amplitude and the period increase up to
$\Delta R/\langle R\rangle\approx 0.6$ and $\Pi=62$ d, respectively, so that stellar oscillations
become non--linear and less regular.
Therefore, the hydrodynamic model shown in Fig.~\ref{fig3} qualitatively reproduces the main
evolutionary features observed in V725 Sgr.

\begin{figure}
\centerline{\includegraphics{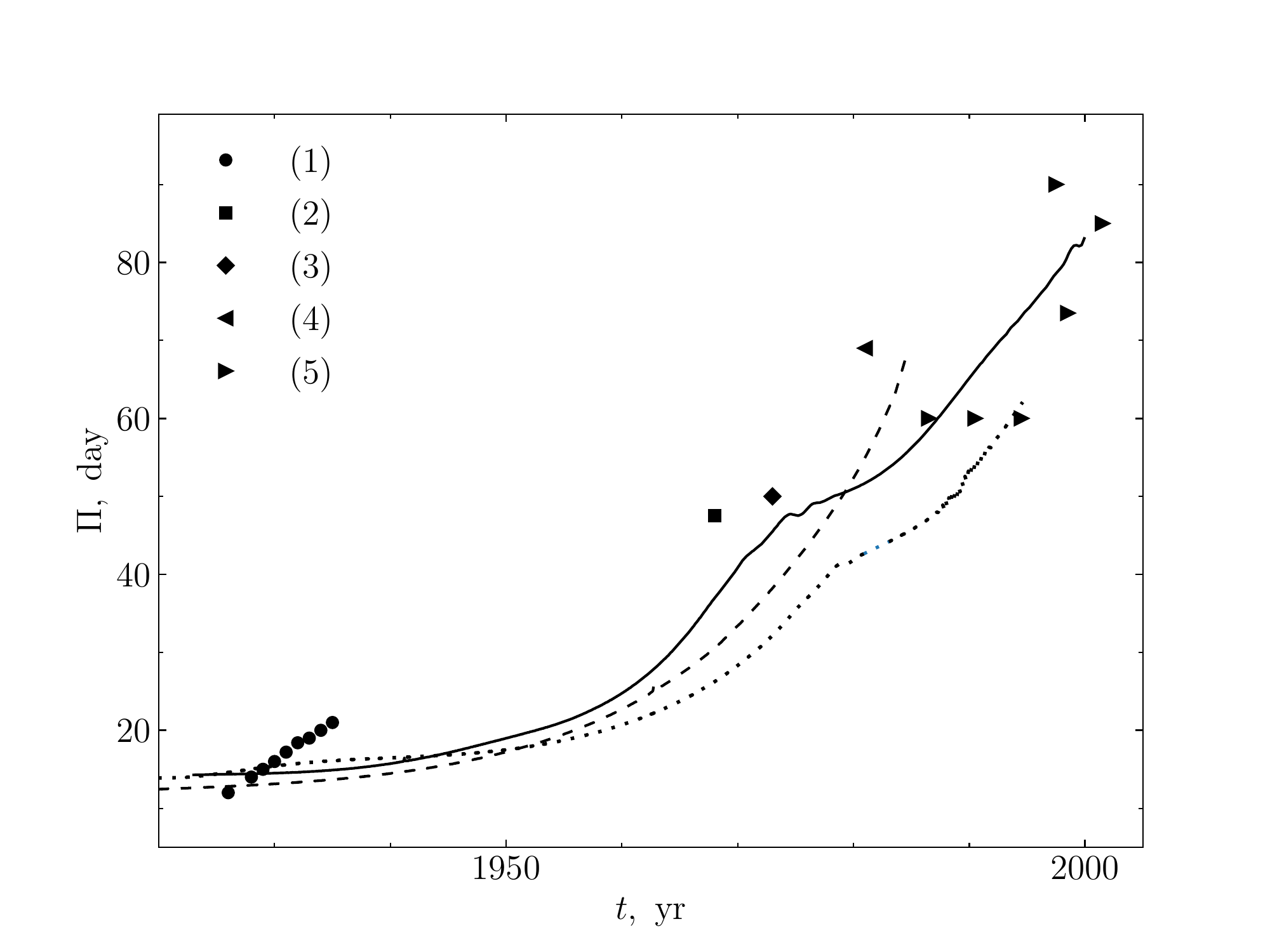}}
\caption{The period of light changes in V725~Sgr from (1) Swope (1937); (2) Demers (1973);
        (3) Demers and Madore (1974); (4) Wehlau et al. (2006); (5) Percy et al. (2006).
        Results of hydrodynamic calculations are shown for evolutionary sequences
        $Z=0.006$, $\mzams=0.88M_\odot$, $\etar=0.5$, $\etab=0.08$ (solid line),
        $Z=0.006$, $\mzams=0.84M_\odot$, $\etar=0.5$, $\etab=0.05$ (dashed line) and
        $Z=0.010$, $\mzams=0.90M_\odot$, $\etar=0.5$, $\etab=0.03$ (dotted line).}
\label{fig4}
\end{figure}

Fig.~\ref{fig4} shows the observational estimates of the pulsation period of V725 Sgr obtained by
Swope (1937), Demers (1973), Demers and Madore (1974), Wehlau e t al. (2006) and Percy et al. (2006).
Theoretical dependences $\Pi(\tev)$ obtained from calculations of three hydrodynamic models
are shown also in Fig.~\ref{fig4}.
It should be noted that for the sake of convenience the plots of $\Pi(\tev)$ are shifted along the
horizontal axis to fit the minimum period to the date $t=1926$ yr when the pulsation period of V725 Sgr
was 12 day.
Main properties of hydrodymanic models shown in Fig.~\ref{fig4} are listed in Table~\ref{tabl1}.

\section*{conclusions}

\begin{figure}
\centerline{\includegraphics{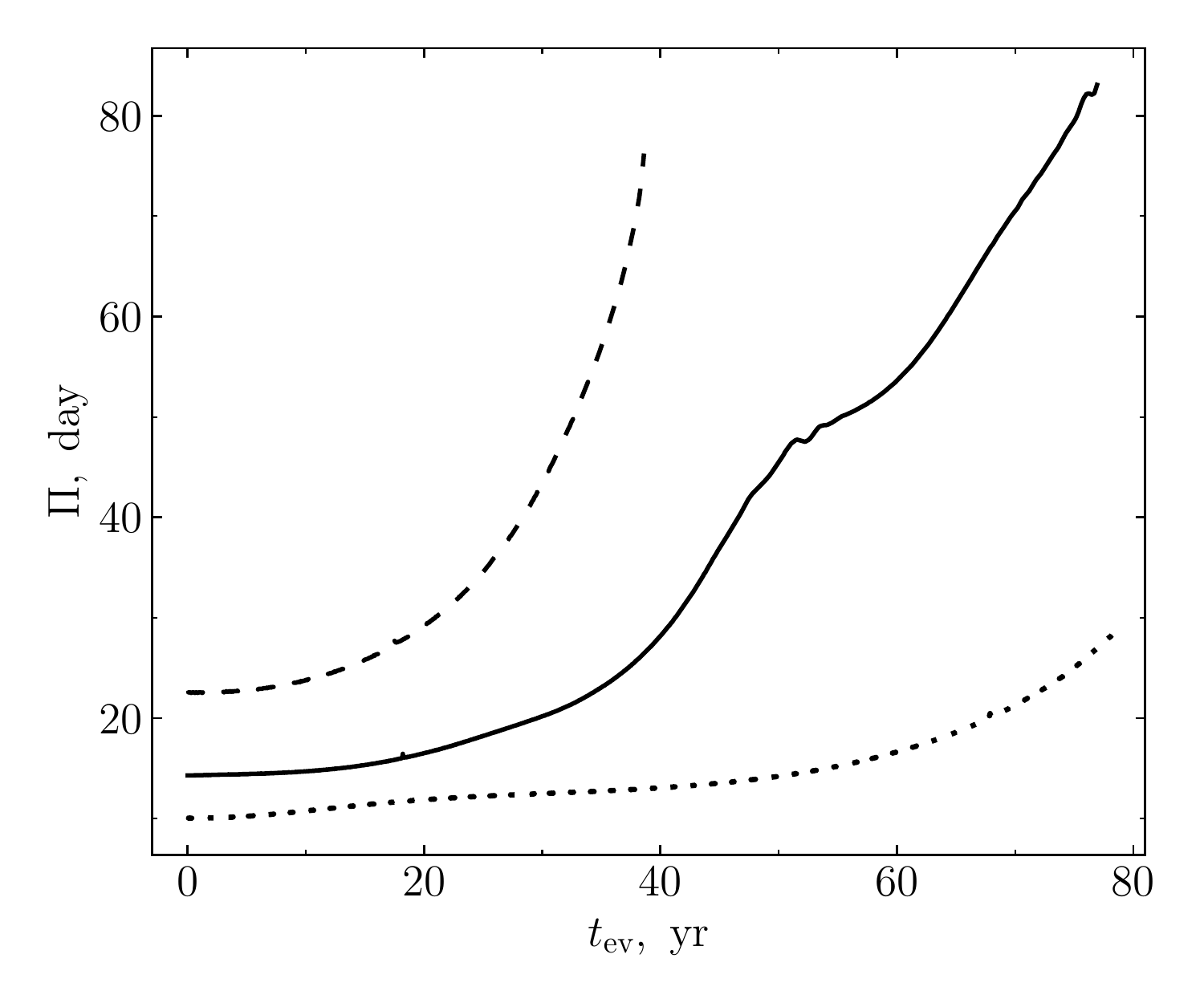}}
\caption{The period of radial pulsations as a function of time for hydrodynamic models
         $Z=0.006$, $\mzams=0.84M_\odot$, $\etar=0.3$, $\etab=0.05$ (dashed line, $\Pi_{\min}=22.5$ day),
         $Z=0.006$, $\mzams=0.88M_\odot$, $\etar=0.5$, $\etab=0.08$ (solid time, $\Pi_{\min}=14.3$ day) and
         $Z=0.006$, $\mzams=0.84M_\odot$, $\etar=0.5$, $\etab=0.06$ (dotted line, $\Pi_{\min}=10.0$ day).
         The evolutionary time is set to zero at the minimum of the stellar radius.}
\label{fig5}
\end{figure}

Results of stellar evolution and non--linear stelar pulsation calculations presented above confirm
the hypothesis by Percy et al. (2006) that the period change observed in V725 Sgr is due to
the thermal flash of the helium burning source of the population II low--mass post--AGB star.
Moreover, a satisfactory agreement between three theoretical dependences $\Pi(\tev)$ and the
observed secular period change in V725 Sgr allowed us to obtain approximate theoretical estimates
of the stellar mass and the mass of the hydrogen envelope: $M\approx 0.53M_\odot$,
$0.013M_\odot\le\mh\le 0.019M_\odot$.
It should be noted that uncertainties in estimates of the stellar mass are due to both
the significant scatter and the small number of observational estimates of the pulsation period
of V725 Sgr.
The minimum value of the pulsation period of V725 Sgr in the beginning of the XX--th century
remains unknown because all available results of observations show the period growth.
Nevertheless, results of our computations allow us to assume that the minimum value of the period
can only slightly differ from the value $\Pi=12$ day.
This is due to the fact that the common feature of hydrodynamic models
is that the characteristic time of period growth increases with increasing $\Pi_{\min}$.
This is illustrated in Fig.~\ref{fig5} where we plot three dependences of period change $\Pi(\tev)$
obtained from computation of hydrodynamic models
$Z=0.006$, $\mzams=0.84M_\odot$, $\etar=0.3$, $\etab=0.05$ ($\Pi_{\min}=22.5$ day),
$Z=0.006$, $\mzams=0.88M_\odot$, $\etar=0.5$, $\etab=0.08$ ($\Pi_{\min}=14.3$ day) and
$Z=0.006$, $\mzams=0.84M_\odot$, $\etar=0.5$, $\etab=0.06$ ($\Pi_{\min}=10.0$ day)
where the only one model (with period $\Pi_{\min}=14.3$ day) agrees well with observations
(see Fig.~\ref{fig4}).
Therefore, it is unlikely that the minimal value of the pulsation period in V725 Sgr was less
than 10 day because in such a case the growth time of the period is too slow.

\newpage
\section*{references}

\begin{enumerate}
\item T. Bl\"ocker, Astron. Astrophys. \textbf{297}, 727 (1995).

\item S. Demers, J. Royal Astron. Soc. of Canada \textbf{67}, 19 (1973).

\item S. Demers and B.F. Madore, Inform. Bull. Var. Stars \textbf{870}, 1 (1974).

\item S. Demers and W.E. Harris, Astron. J. \textbf{79}, 627 (1974).

\item Yu.A. Fadeyev, Astron. Lett. \textbf{39}, 306 (2013).

\item Yu.A. Fadeyev, Astron. Lett. \textbf{46}, 734 (2020).

\item Yu.A. Fadeyev, Astron. Lett. \textbf{47}, 765 (2021).

\item Yu.A. Fadeyev, MNRAS \textbf{514}, 5996 (2022).

\item H.C. Harris, Astron. J. \textbf{86}, 719 (1981).

\item H.C. Harris and G. Wallerstein, Astron. J. \textbf{89}, 379 (1984).

\item B. Paxton, R. Smolec, J. Schwab, A. Gautschy, L. Bildsten, M. Cantiello,
      A. Dotter,  R. Farmer, J.A. Goldberg, A.S. Jermyn, S.M. Kanbur, P. Marchant, A. Thoul,
      R.H.D. Townsend, W.M. Wolf, M. Zhang, and F.X. Timmes,
      Astrophys. J. Suppl. Ser. \textbf{243}, 10 (2019).

\item J.R. Percy, J. Am. Associat. Var. Star Observ. \textbf{48}, 162 (2020).

\item J.R. Percy, A. Molak, H. Lund, D. Overbeek, A.F. Wehlau and P.F. Williams,
      Publ. Astron. Soc. Pacific \textbf{118}, 805 (2006).

\item D. Reimers, \textit{Problems in stellar atmospheres and envelopes}
      (Ed. B. Baschek, W.H. Kegel, G. Traving, New York: Springer-Verlag, 1975), p. 229.

\item N.N. Samus', E.V. Kazarovets, O.V. Durlevich, N.N. Kireeva, and E.N. Pastukhova,
      Astron. Rep. \textbf{61}, 80 (2017).

\item H.H. Swope, Annals of Harvard College Observatory \textbf{105}, 499 (1937).

\item E. Vassiliadis and P.R. Wood, Astrophys. J. \textbf{413}, 641 (1993).

\item A. Wehlau, T. Atcheson T. and S. Demers,
      J. Am. Associat. Var. Star Observ. \textbf{35}, 187 (2006).

\end{enumerate}

\newpage
\begin{table}
\caption{Hydrodynamic models V725 Sgr}
\label{tabl1}
\begin{center}
\begin{tabular}{c|c|c|c|c|c|c}
\hline
$Z$ & $\mzams/M_\odot$ & $\etar$ & $\etab$ & $M/M_\odot$ & $\mh/M$ & $\Pi_{\min}$, d \\
\hline
0.006 & 0.84 & 0.5 & 0.05 & 0.527 & 0.037 & 11.3 \\
0.006 & 0.88 & 0.5 & 0.08 & 0.534 & 0.025 & 14.2 \\
0.010 & 0.90 & 0.5 & 0.03 & 0.534 & 0.020 & 13.9 \\
\hline
\end{tabular}
\end{center}
\end{table}
\clearpage

\newpage
\clearpage

\newpage
\clearpage

\newpage
\clearpage

\newpage
\clearpage

\newpage
\clearpage

\end{document}